\documentclass[preprint,12pt]{elsarticle}

\usepackage{amssymb}
\usepackage{float}

\journal{Arxiv}

\begin{document}

\begin{frontmatter}



\title{Evaluation of patient activation and dosimetry after Boron Neutron Capture Therapy}


\author[inst1,inst2]{Giovanni Garini}

\affiliation[inst1]{organization={University of Pavia, Department of Physics},
            addressline={via A. Bassi 6}, 
            city={Pavia},
            postcode={27100}, 
            state={Italy},
            }
            
 \affiliation[inst2]{organization={National Institute for Nuclear Physics (INFN), Unit of Pavia},
            addressline={via A. Bassi 6}, 
            city={Pavia},
            postcode={27100}, 
            state={Italy},
            }

\affiliation[inst3]{organization={Emme Esse Srl},
            addressline={via privata Giuba 11}, 
            city={Milan},
            postcode={20132}, 
            state={Italy},
            }

           \affiliation[inst4]{organization={University of Campania "Luigi Vanvitelli", Department of Mathematics and Physics},
           addressline={Viale Abramo Lincoln 5}, 
           city={Caserta},
           postcode={81100}, 
          state={Italy},
           }

            \affiliation[inst5]{organization={University of Campania "Luigi Vanvitelli", Department of Architecture and Industrial Design},
           addressline={Viale Abramo Lincoln 5}, 
           city={Caserta},
           postcode={81100}, 
          state={Italy},
           }

             \affiliation[inst6]{organization={University of Campania "Luigi Vanvitelli", Department of Advanced Surgical Medical Sciences},
           addressline={Viale Abramo Lincoln 5}, 
           city={Caserta},
           postcode={81100}, 
          state={Italy},
           }
\author[inst3]{Chiara Magni}
\author[inst2]{Ian Postuma}
\author[inst2]{Setareh Fatemi}
\author[inst2]{Ricardo Ramos}
\author[inst1, inst2]{Barbara Marcaccio}
\author[inst1, inst2]{Cristina Pezzi}
\author[inst4]{Laura Bagnale}
\author[inst5]{Sandro Sandri}
\author[inst5]{Gianfranco De Matteis}
\author[inst6]{Giuseppe Paolisso}
\author[inst2]{Valerio Vercesi}
\author[inst1,inst2]{Silva Bortolussi*}

\begin{abstract}
Boron Neutron Capture Therapy (BNCT) is a form of radiotherapy based on the irradiation of the tumour with a low energy neutron beam, after the administration of a selective drug enriched in boron-10. The therapy exploits the high cross section of thermal neutron capture in boron, generating two low-range charged particles. The availability of accelerators able to generate high-intensity neutron beams via proton nuclear interaction is boosting the construction of new clinical centres. One of these is under development in Italy, using a 5 MeV, 30 mA proton radiofrequency accelerator coupled to a beryllium target, funded by the Complementary Plan to the Recovery and Resilience National Plan, under the project ANTHEM. The present study focuses on radiation protection aspects of patients undergoing BNCT, specifically on the activation of their organs and tissues. A criterion to establish the relevance of such activation after BNCT has been proposed. Based on the current Italian regulatory framework, the level of patient activation following BNCT treatment does not pose a significant radiological concern, even shortly after irradiation. Another aspect is the activation of patient's excretions, which can impact on the design of the building and requires a process for the discharge. The described study contributes to the radiation protection study for the ANTHEM BNCT centre in Italy.
\end{abstract}


\begin{highlights}
\item The adult-patient model from ICRP-145 has been used for a benchmark between MCNP and PHITS
\item A criterion for patients discharge after BNCT based on Italian regulation in nuclear medicine is proposed
\item Activation of patients after BNCT has been studied in different irradiation positions 
\item Excretes activation has been evaluated
\item Activation studies are pivotal in the BNCT facility design and management

\end{highlights}


\end{frontmatter}


\section{Introduction}
\label{sec:intro}
The project ANTHEM (AdvaNced Technologies for Human-centrEd Medicine) is funded by the National Plan for Complementary Investments (PNC) in the call for technologies and innovative trajectories in the health and care sectors.
Its purpose is to fill the existing gap in the healthcare of fragile and chronic patients within specific territories, characterized by pathologies that lack effective therapies. \newline
The project is organized into four spokes with different aims;
the Pilot 9 of the Spoke 4 concerns the construction of a new facility for Boron Neutron Capture Therapy (BNCT) research and clinical application in Caserta, at the University of Campania "Luigi Vanvitelli", Italy. 

BNCT is a form of radiotherapy consisting in the irradiation with low energy neutrons of a tumour previously loaded with a sufficient concentration of boron-10 nuclei. The cross section of thermal neutron capture in$^{10}$B is orders of magnitude higher than that for interaction with other elements in biological tissues. The reaction $^{10}$B(n,$\alpha$)$^{7}$Li produces two particles with high Linear Energy Transfer (LET) and short range in biological tissue, which cause lethal damage to the cell where the capture occurs. Thus, provided a suitable tumour-to-normal tissue boron concentration ratio is achieved, a differential radiation dose deposition is attained in the malignancy, with a substantial sparing of healthy tissues \cite{IAEA-TecDoc2023}.

The technology to generate the neutron beam for the ANTHEM project is designed and built by National Institute for Nuclear Physics (INFN) and it is based on a proton accelerator, a Be target and a Beam Shaping Assembly (BSA). The accelerator, developed by the National Laboratory of Legnaro (LNL) is a Radio Frequency Quadrupole (RFQ) machine delivering a 5 Mev, 30 mA proton beam in continuous wave \cite{pisent2014munes}. The proton beam hits a Be-Pd-Cu target, also designed at LNL. Finally, the BSA to tailor the energy spectrum and collimation for patients is being developed in Pavia using an innovative method to select the best configuration of materials and to evaluate the therapeutic potential of the beam \cite{postuma2021novel}. 
The neutron beam will be characterized by a flux of the order of $10^{9} cm^{-2} s^{-1}$, as required by the IAEA guidelines \cite{IAEA-TecDoc2023}, starting from a neutron intensity of  $10^{14} s^{-1}$ produced in the target \cite{ESPOSITO}.

\begin{figure}[h!]
\centering
\includegraphics[width=14.5cm]{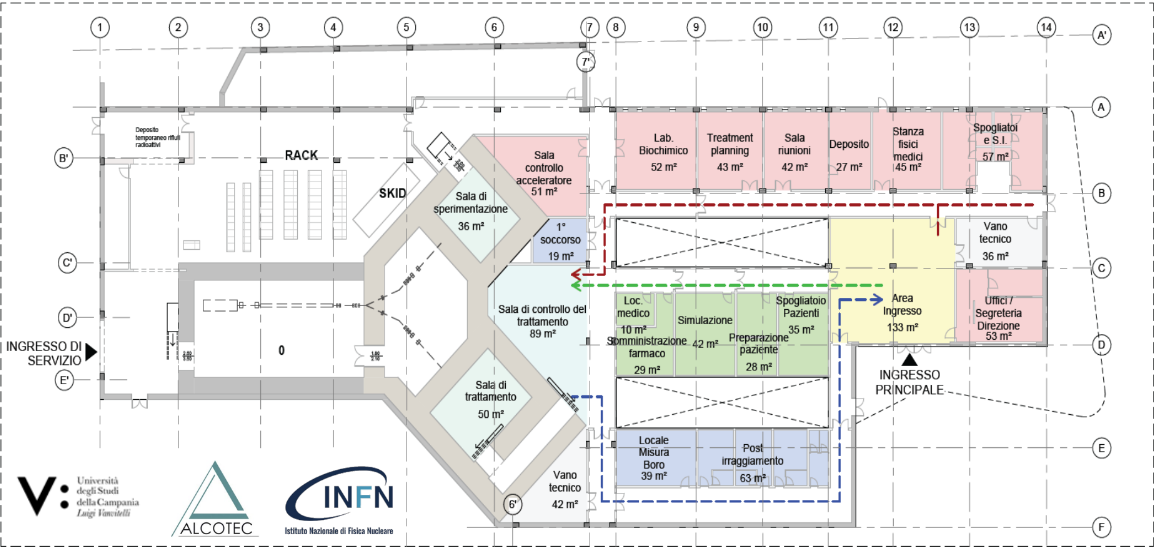}
\caption{Building plan of ANTHEM project facility for BNCT in Caserta, Italy.}
\label{images}
\end{figure}

The Pilot 4.9 of ANTHEM also addresses the design and construction of the building comprising the spaces for the neutron production technology, one patients' irradiation room, one experimental room, and areas for patients and medical/research staff as shown in the plan in Fig. \ref{images}.

The building will include rooms dedicated to the staff and patient preparation (boron administration and positioning), two irradiation rooms, one for clinics and one for research purposes, and laboratories for radiobiology, biochemistry and molecular medicine. 



The irradiation of patients with an intense neutron beam generates induced activation in the biological tissues and in the excretions and may pose some constraints in the management of patients in the post-treatment phase, an aspect to be addressed in accordance with the legislation currently in force in Italy regarding radiation protection. 
Patients activation could be a safety issue for people who will come in contact with them after the treatment, such as medical staff and family members. For this reason, it is necessary to evaluate the patient induced radioactivity and to calculate the dose that a person will absorb by being in proximity of the patient. 
This work presents these calculations and proposes a criterion to evaluate the dose due to this activation to take decisions on the discharge of patients, as there is no specific mention of this aspect in the radiation protection legislation concerning BNCT or neutron irradiation. 
The obtained results contribute to the radiation protection studies connected to the approval of the BNCT centre, to the design of the facility and to the establishment of the procedures from the patients admission to their discharge after the irradiation.  

\section{Materials and methods}
The current legislation governing radiation protection in Italy is Legislative Decree 101/2020, whose Title XIII contains the provisions on medical exposure to ionizing radiation.
In accordance with the principles of radiation protection, in the case of patients treated with radiopharmaceuticals it is necessary to protect the people who will come in contact with them after treatment.
As the Decree states in Annex XXV, post-treatment hospitalization is mandatory only for patients administered with Iodine-131 ($^{131}I$) at activity levels greater than 600 MBq. 
It can reasonably be assumed that, below this threshold, contact with a treated patient is safely acceptable from a radiation protection viewpoint.
Therefore, since there are currently no specific provisions for patients undergoing neutron irradiation, we decided to compare the ambient dose due to the activation of a patient treated with BNCT to that of a patient treated with 600 MBq of $^{131}I$.
As the quantity to calculate, we chose the ambient dose, a radiation protection parameter used to assess the dose in the air within a specific environment and the associated risk for individuals in that area. 

\subsection{ICRP phantom}

\begin{figure}
\centering
 [a]
{\includegraphics[width=4.6cm]{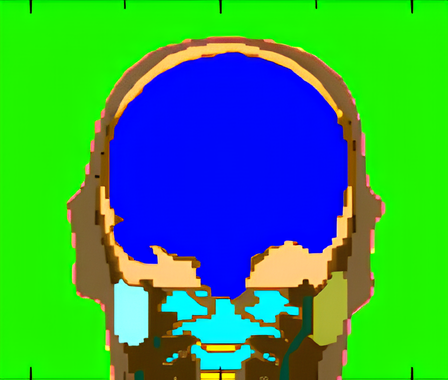}}
\hspace{0.1mm}
 [b]
{\includegraphics[width=3.2cm]{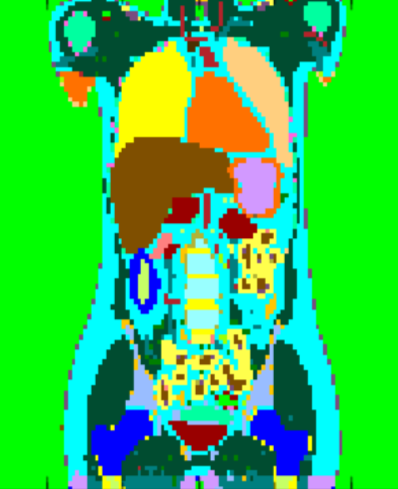}}   
\hspace{0.1mm}
 [c]
{\includegraphics[width=2cm]{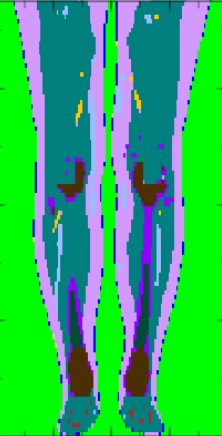}} 
\caption{Phantom particulars of: [a] head district,[b] thoracic district, [c] lower limbs.}
    \label{phant}
\end{figure}

To simulate $^{131}I$ treatment and BNCT irradiation, we selected 
the new adult mesh-type computational phantom from the ICRP publication nr. 145 \cite{icrp145},
made with a tetrahedral mesh which offers a high resolution, especially for small structures.
Figure \ref{phant} shows the three districts of the computational phantom that were considered for the BNCT irradiation in this paper.

We compared MCNP6.3 \cite{MCNP6.3}, which is a gold standard in BNCT computational dosimetry, and PHITS3.33 \cite{PHITS}, which offers some advantages for the purposes of this work. We performed a benchmark simulation to compare the results obtained by the two codes and evaluated the computation time.
The machine used for calculations has two CPUs (Intel(R) Xeon(R) CPU E5-2680 v3 @ 2.50GHz), with 24 threads, and 64 GB of RAM each. To display the geometry, PHITS needs a few minutes, while MCNP needs some days.
Moreover, due to the type of mesh implemented in MCNP, parallel calculation is not possible, leading to a very long computing time. PHITS, on the other hand, can implement a parallel calculation with this geometrical model. 

PHITS allows the use of the code DCHAIN-SP, taking as input the output of the PHITS calculation of neutron activation and generating a source for the calculation of the dose due to the decay of the activated species \cite{RATLIFF}. In the PHITS input, one must specify 
the parameters of the regions for which induced activity is requested, i.e., how long the source is on and its intensity, and 
how much time must elapse between the shutdown of the beam and the evaluation of the activation.
This produces, among other results, also the definition of a source with the spectrum of the isotopes activated in the selected regions, that can be used as the source for a new PHITS simulation.


The benchmark simulation consisted in the calculation of the dose in air due to 600 MBq of $^{131}I$ administered to the patient for radionuclide metabolic therapy, that we assumed concentrated in the thyroid. The photon flux and the ambient dose were evaluated around the patient in a cylindrical mesh, with axis along the vertical axis of the phantom (z-axis), scoring the upper part of the phantom (from z=0 to z=80 cm) and radially from r=0 to r=120 cm in steps of 10 cm.
As a figure of merit, we report the relative difference (RD) as:

\begin{equation} \label{eq_1}
RD= \frac{|\phi_{MNCP}-\phi_{PHITS}|}{\phi_{MNCP}}.   
\end{equation} 

To convert flux into ambient dose H* we used tabulated factors from the latest ICRU release nr. 95 \cite{ICRU95} .
The RD was calculated by substituting the ambient dose values for the fluxes in equation \ref{eq_1}.
After the evaluation of the RD in the benchmark simulation, we used PHITS for the rest of the calculations. 

\subsection{Phantom positioning and irradiation setup}
\begin{figure}
\centering
 [a]
{\includegraphics[width=6.25cm]{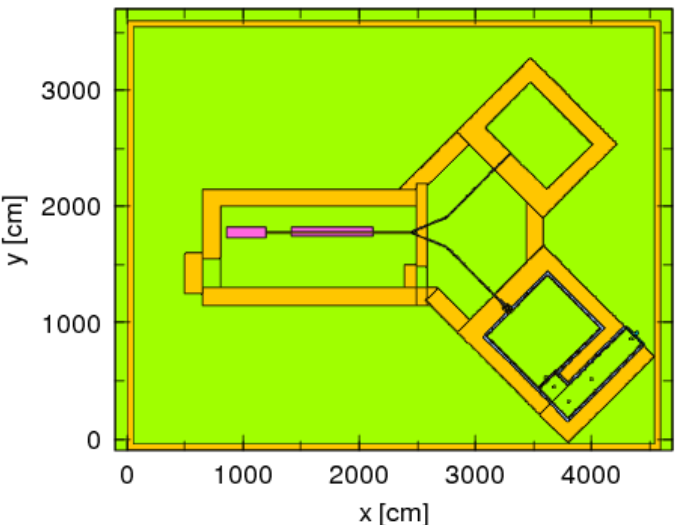}}
\hspace{5mm}
 [b]
{\includegraphics[width=5.5cm]{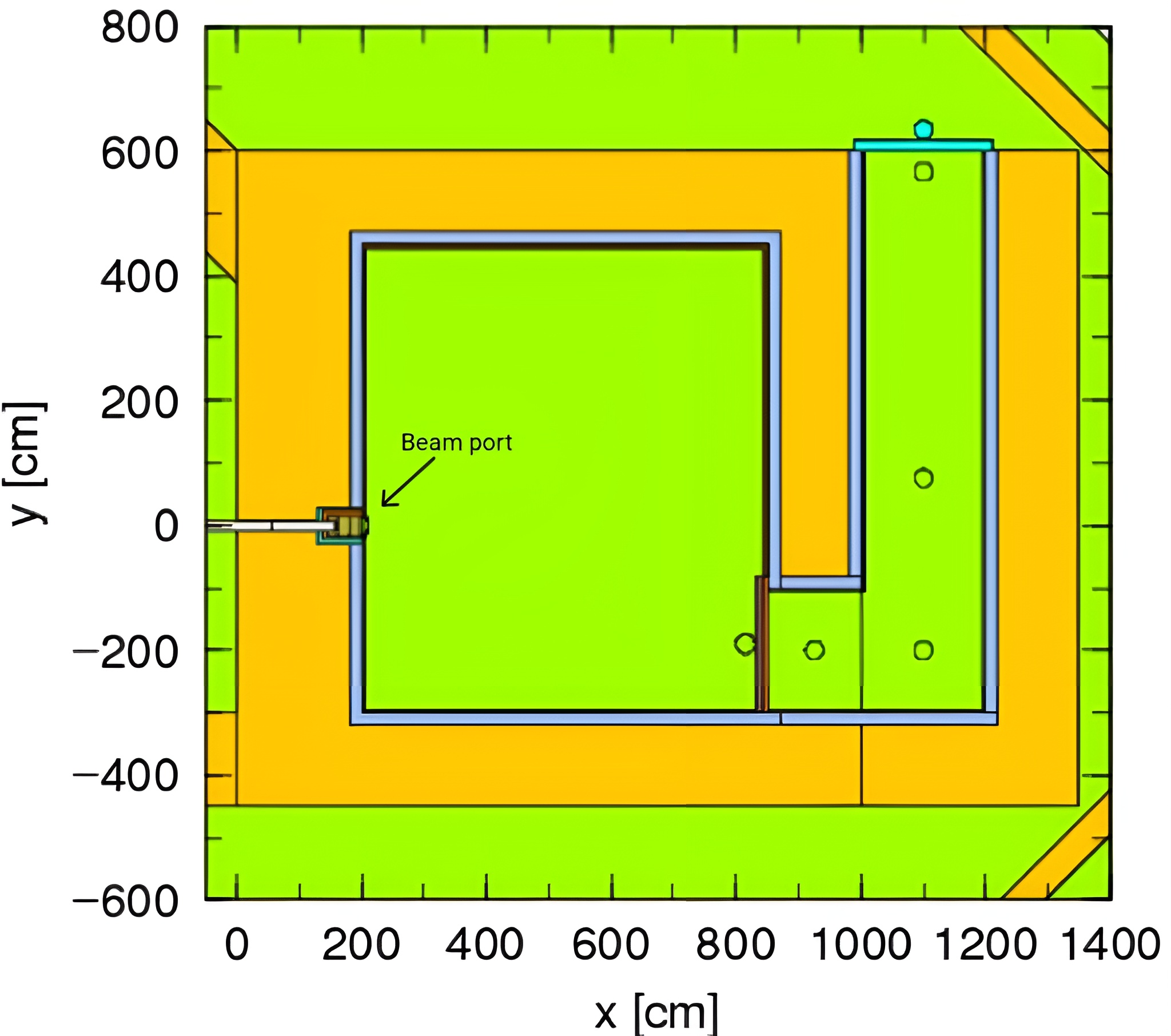}}   
\caption{Facility geometry in PHITS [a] and a particular of the treatment room [b].}
    \label{image6}
\end{figure}

To obtain realistic results for patient activation due to BNCT, the geometry of the room was reproduced to consider all the relevant neutron interactions with the walls and with the patient (Figure \ref{image6}). 

\begin{figure}
\centering
{\includegraphics[width=13.5 cm]{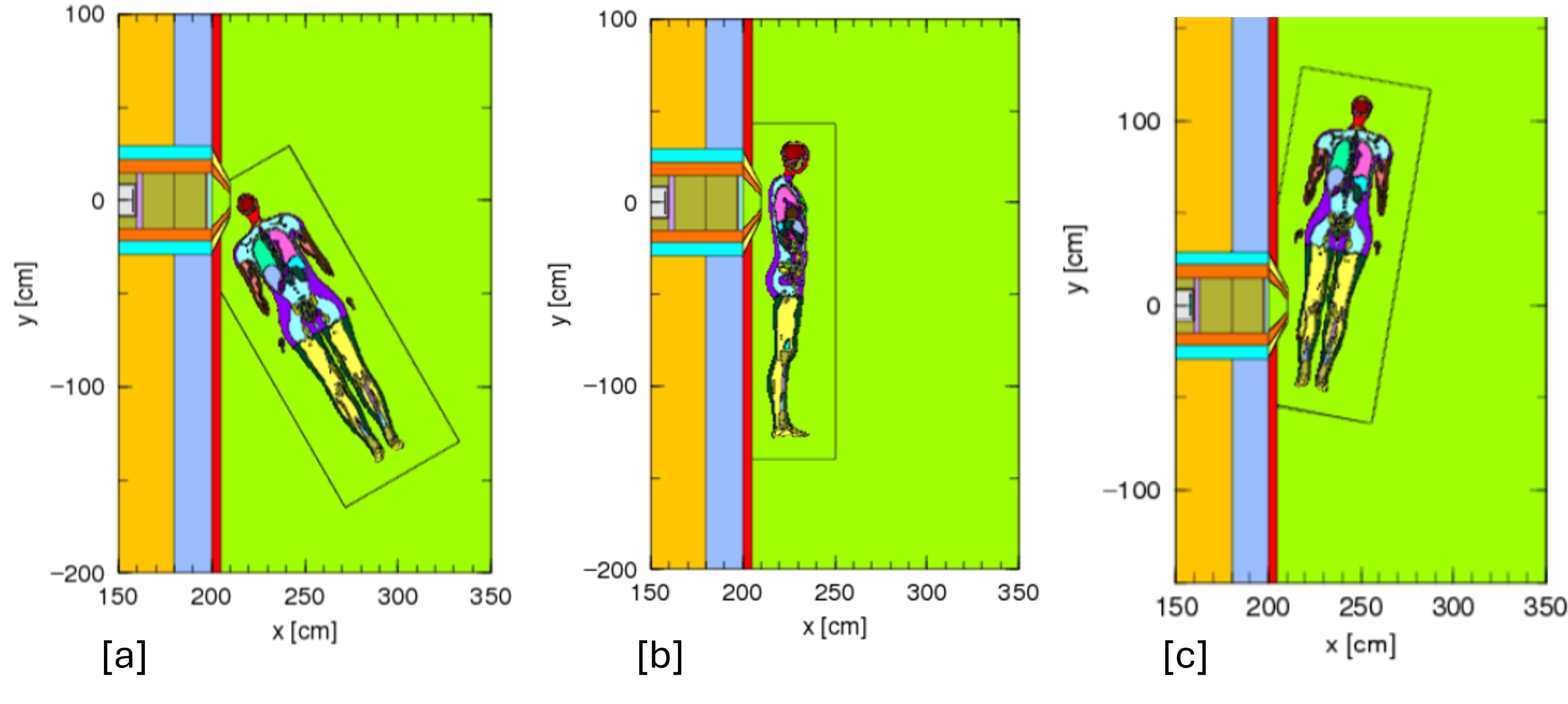}}

    \caption{Patient irradiation positions for cancers located in: neck-head region [a], thorax [b], lower limbs [c].}
    \label{image7}
\end{figure}

The wall composition had been previously optimized for shielding the radiation 
as a three-layer structure, made up of a layer of Portland concrete (130 cm thick), a layer of baritic concrete (20 cm thick), and a layer of borated polyethylene (5 cm thick)  working as a neutron absorber. 
Patient activation has been evaluated in three different irradiation positions simulating the treatment of cancers located in the head-neck region (Fig.\ref{image7}[a]), in the thorax (Fig.\ref{image7}[b]) and in the lower limbs (Fig.\ref{image7}[c]).

These positions are only representative for studying the activation.
The spectrum of the neutron beam irradiating the patient is described in \cite{postuma2021novel}. 

The number of neutrons for each simulation has been chosen evaluating the uncertainty in each cell composing the phantom when tallying neutron flux, with a \textbf{[T-Track]}. In order to avoid an increase of the calculation time, we fixed as an acceptable threshold a relative error lower than 15\% in cells where the flux is, at most, two orders of magnitude lower than the maximum value.
This required to simulate, for each irradiation position, 100 batches of $10^6$ neutrons each, to obtain statistically significant results.

To obtain information about patient activation, a \textbf{[T-Dchain]} tally was used in each region of the phantom and considering 1 and 2 hours of irradiation time with two different tallies.
For each of these tallies five output times have been considered, one at the end of irradiation (0 s) and at 10, 15, 30, 45 minutes after the beam shutdown.

\subsection{Excretions}
For the executive plan of the facility and in view of its commissioning, the evaluation of the activation of urine helps in planning their disposal.
The activated excretions must be in fact collected in special shielded containers waiting for the activity reduction under a certain threshold.
The bladder urinary content provided in the phantom from \cite{icrp145} - cell (N° 13800) - lacks of some elements like chlorine and sulfur that are relevant for activation analysis. The latter is significant because of the reaction $^{34}$S(n,$\gamma$)$^{35}$S and the following production of $^{35}$S with a half-life of 87.5 days \cite{magni2022experimental}.
For this reason the urine composition has been changed 
according to \cite{woodard1986composition} as reported in Table \ref{urine}.

\begin{table}[H]
\begin{tabular}{c|c|c|c|c|c|c|c|c|c}
  & H & C & N & O & Na & P & S & Cl & K \\ \hline
Default & 10.7 & 0.3 & 1 & 87.1 & 0.4 & 0.1 & - & - & 0.2 \\ \hline
New & 10.978 & 0.499 & 0.988 & 86.0276 & 0.3992 & 0.998 & 0.2 & 0.5988 & 0.1996
\end{tabular}
\caption{Urine composition in percentage.}
\label{urine}
\end{table}

\section{Results}\label{sec:res}

\subsection{PHITS validation against MCNP}
The RD factor has been calculated for each zone of the cylindrical mesh around the patient model for the benchmark evaluation of PHITS compared to MCNP. The obtained values are all below 9\% and in 68.5\% of the cases the results are below 5\%, thus showing a good agreement. The relative difference can be explained considering that MCNP and PHITS use different data libraries.
The dose difference in percentage was evaluated with the same formula used for the flux. For all the mesh regions except one, the values of the relative difference of dose were below  5\%. 

Table \ref{table1} reports  the comparison of the total flux in the cylinder for the two codes and the total H*.

\begin{table}
\centering
\begin{tabular}{l|l|l|l}
& MCNP & PHITS  & Variation (\%) \\ \hline
Flux ($cm^{-2} \cdot s^{-1}$)  & 5.20$\cdot 10^{6}$ $\pm$ 3.17$\cdot 10^{3}$
 & 5.56$\cdot 10^{6}$
$\pm$ 2.42$\cdot 10^{4}$
 & \multicolumn{1}{c}{6.9}
\\ \hline
\hline
& MCNP & PHITS  & Variation (\%) \\ \hline
H* (pSv $\cdot s^{-1}$) & 7.00$\cdot 10^{6}$ $\pm$ 4.15$\cdot 10^{3}$
 & 7.16$\cdot 10^{6}$ $\pm$ 3.19$\cdot 10^{4}$
 & \multicolumn{1}{c}{2.3}

\end{tabular}
\caption{Comparison of MCNP and PHITS flux and H* calculations}
\label{table1}
\end{table}



The discrepancy between the dose values is lower than the one obtained in the flux calculations, to be possibly attributed to the logarithmic interpolation which compensates for the flux difference.

Given these results, PHITS was adopted for the following evaluations.

\subsection{$^{131}$I in thyroid}
\begin{figure}[H]
    \centering
    \includegraphics[width=9cm]{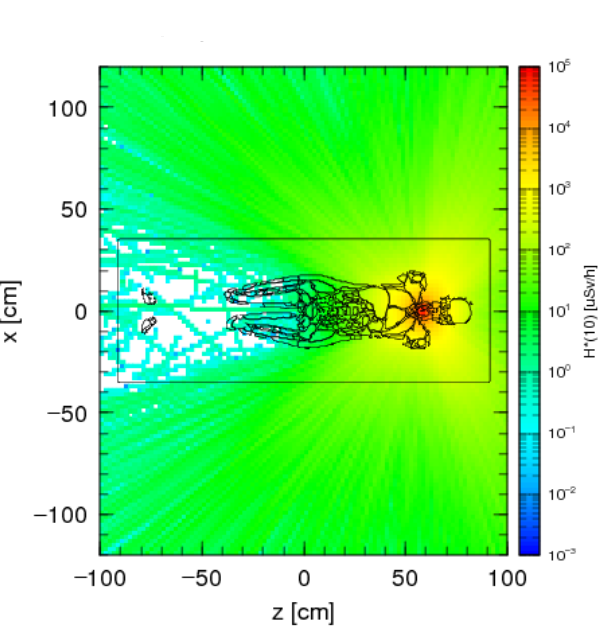}
    \caption{H* rate in a radius of 1.2 m around the patient due to 600 MBq of $^{131}$I in the thyroid.}
    \label{image5}
\end{figure}

Figure \ref{image5} shows the results of the ambient dose calculation in PHITS with a larger mesh going from z=-80 cm and z=80 cm (whole body height) that was useful to define the limit for the patient discharge.

\begin{table}[H]
\centering
\caption{H* rate around the patient due to 600 MBq of $^{131}$I in the thyroid.} 
\label{tableH10Iodio}
\begin{tabular}{c|c|c}
 Radius (cm)  &  H* 1h irr. ($\mu$Sv/h)  & Rel. err. \\ \hline
30-40 & 1.15$\cdot 10^{3}$ & 7.61$\cdot 10^{-3}$ \\ \hline
40-50 & 8.18$\cdot 10^{2}$ & 8.56$\cdot 10^{-3}$ \\ \hline
50-60 & 6.22$\cdot 10^{2}$ & 9.28$\cdot 10^{-3}$ \\ \hline
60-70 & 4.94$\cdot 10^{2}$ & 9.91$\cdot 10^{-3}$ \\ \hline
70-80 & 4.05$\cdot 10^{2}$ & 1.04$\cdot 10^{-2}$ \\ \hline
80-90 & 3.40$\cdot 10^{2}$ & 1.09$\cdot 10^{-2}$ \\ \hline
90-100 & 2.89$\cdot 10^{2}$ & 1.12$\cdot 10^{-2}$\\ \hline
100-110 & 2.49$\cdot 10^{2}$ & 1.16$\cdot 10^{-2}$ \\ \hline
110-120 & 2.17$\cdot 10^{2}$ & 1.19$\cdot 10^{-2}$\\ 
\end{tabular}
\end{table}

Table \ref{tableH10Iodio} shows the values of H* rate around the patient from a distance of 30 cm to 120 cm due to an activity of 600 MBq of $^{131}$I placed in the thyroid. These values 
were the reference to understand the possibility of patient discharge after BNCT treatment. 

\subsection{Activation of Patient}
\subsubsection{Irradiation of the head and neck district}
When the patient is positioned for irradiation of the region of the head and neck, the three most activated regions, at any irradiation time, are:
\begin{itemize}
    \item N° 2600, the cranium cortical bone;
    \item N° 2700, the cranium spongiosa bone;
    \item N° 6100, the brain. 
\end{itemize}
Tables \ref{table2.2}, 
\ref{table2.4}, 
\ref{table2.6},
list, for the three regions, the three most relevant isotopes, in terms of relative percentage of total activation, for 1 hour of irradiation, at the end of the treatment (+0 s) and 15 minutes after the beam shutdown (+ 15 m). 
The activity reported is the sum of the contributions from all the decay types (beta and gamma), although only the gamma emission plays a role in the calculation of ambient dosimetry.

\begin{table}[H]
\centering
\caption{Activation of isotopes in Region 6100 (brain), for 1 h Irradiation, at the beam shutdown (+ 0 s) and after 15 minutes (+15 m). Statistical uncertainty below  0.5\%.} 
\label{table2.2}
\begin{tabular}{c|c|c|c|c|c}
 Isotope      & Activity(Bq)  & Rel activity &  Activity(Bq) & Rel activity &  Half-life (s) \\ 
  & +0 s & +0 s & +15 m & +15 m & \\ \hline
 $^{24^m}$Na & 1.982 $\cdot 10^7$ & 80.29\%  & - & -  & 2.018 $\cdot 10^{-2}$ \\ \hline
  $^{38}$Cl & 3.401 $\cdot 10^6$ & 13.78\%   & 2.572 $\cdot 10^6$ & 64.03\% & 2.234 $\cdot 10^3$ \\ \hline
  $^{24}$Na & 1.192 $\cdot 10^6$  & 4.83\%  & 1.179 $\cdot 10^6$ & 29.34\%   &   5.398 $\cdot 10^4$\\ \hline
   $^{42}$K & - & - & 2.347 $\cdot 10^5$  & 5.84\%   &  4.450 $\cdot 10^4$ 
\end{tabular}
\end{table}


\begin{table}[H]
\centering
\caption{Activation of isotopes in Region 2600 (cranium cortical bone), at the beam shutdown (+ 0 s) and after 15 minutes (+15 m). Statistical uncertainty below  0.5\%.} 
\label{table2.4}
\begin{tabular}{c|c|c|c|c|c}
 Isotope      & Activity(Bq)  & Rel activity &  Activity(Bq) & Rel activity &  Half-life (s) \\ 
  & +0 s & +0 s & +15 m & +15 m & \\ \hline
 $^{24^m}$Na & 4.917 $\cdot 10^6$ & 72.44\% & - & -  & 2.018  $\cdot 10^{-2}$ \\ \hline
  $^{49}$Ca & 1.004 $\cdot 10^6$ & 14.79\% & 3.046 $\cdot 10^5$ & 26.07\%  & 5.231 $\cdot 10^2$ \\ \hline
  $^{49}$Sc & 4.368 $\cdot 10^5$  & 6.44\%   & 4.598 $\cdot 10^5$ & 39.36\%  &  3.431 $\cdot 10^3$ \\ \hline
  $^{24}$Na & - & - & 2.925 $\cdot 10^5$  & 25.03\%   &  5.398 $\cdot 10^4$
\end{tabular}
\end{table}


\begin{table}[H]
\centering
\caption{Activation of isotopes in Region 2700 (cranium spongiosa bone), for 1 h Irradiation, at the beam shutdown (+ 0 s) and after 15 minutes (+15 m). Statistical uncertainty below  0.5\%.} 
\label{table2.6}
\begin{tabular}{c|c|c|c|c|c}
 Isotope      & Activity(Bq)  & Rel activity &  Activity(Bq) & Rel activity &  Half-life (s) \\ 
  & +0 s & +0 s & +15 m & +15 m & \\ \hline
 $^{24^m}$Na & 3.350 $\cdot 10^6$ & 75.68\% & - & -  & 2.018 $\cdot 10^{-2}$ \\ \hline
  $^{49}$Ca & 4.257 $\cdot 10^5$ & 9.62\%  & - & -  & 5.231 $\cdot 10^2$ \\ \hline
  $^{24}$Na & 2.016 $\cdot 10^5$  & 4.55\% & 1.993 $\cdot 10^5$ & 27.32\%   &  5.398 $\cdot 10^4$ \\ \hline
    $^{49}$Sc & - & - & 1.950 $\cdot 10^5$ & 26.74\%  & 3.431 $\cdot 10^3$ \\ \hline
  $^{38}$Cl  & - & -  & 1.438 $\cdot 10^5$  & 19.72\%   & 2.234 $\cdot 10^3$
\end{tabular}
\end{table}


In the three regions, at the beam shutdown the main contribution to the activity comes from $^{24^m}$Na. It decays in $^{24}$Na with a gamma emission with a branching ratio of 99.95\% and in $^{24}$Mg (stable) trough a $\beta ^-$ decay with a branching ratio of 0.05\%.  
Due to the low half-life of this isotope ($\tau$=2.018 $\cdot 10^{-2}$ s) at 15 minutes after the shutdown its contribution is almost zero.
It is also interesting to observe that the longest half lives are of the order of $10^4$ s (some hours). This means that after one day the activity will be reduced by almost one order of magnitude.
The fact that isotopes with half-life in the order of some minutes give the highest contribution to the total activity, suggests that the discharge of persons after a reasonable time post-irradiation is feasible.

\begin{figure}
\centering
 [a]
{\includegraphics[width=5.9cm]{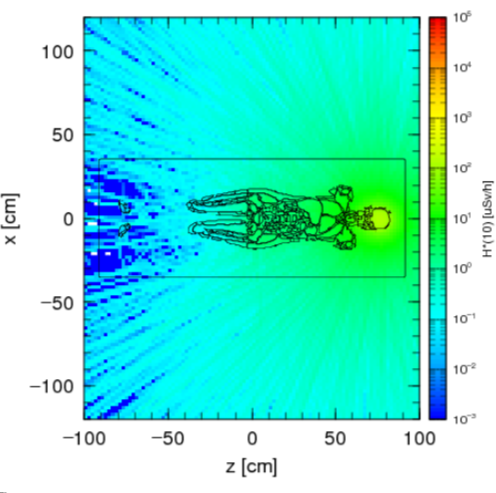}}
\hspace{5mm}
 [b]
{\includegraphics[width=5.8cm]{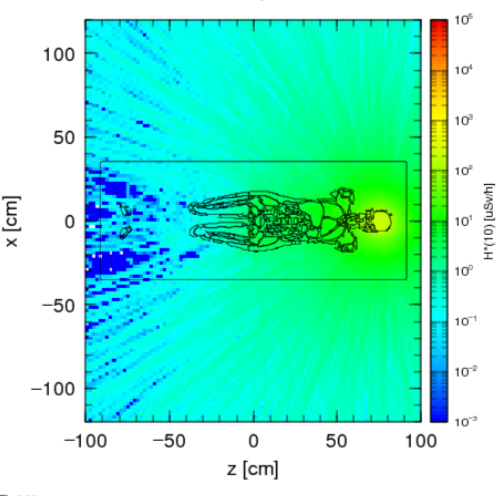}}   
\caption{H* maps at 15 minutes after 1 hour [a] and 2 hours [b] of irradiation in the  head and neck region.}
\label{image9}
\end{figure}

Figure \ref{image9} represents the map of H* rate for 1 and 2 hours of irradiation at 15 minutes after the shutdown. This is a relevant observation time, as patients will likely stay in the irradiation room for some minutes after beam shutdown to allow for the ambient radioactivity to cool down before medical staff can enter.
One can observe that the dose is higher near the head, as expected because the most activated regions are those directly irradiated by the neutron beam.

Note that the H* values are only valid in air around the patient, thus dose values depicted in correspondence of the tissues (between 0 and 30 cm from the axis) are not meaningful.

Table \ref{tableH10head} reports the H* results for both the irradiation times at 15 minutes after the beam shutdown. These values are obtained summing up H* over the whole height of the mesh for each ring of the cylindrical mesh.

\begin{table}[H]
\centering
\caption{H* rate produced by the patient activation due to irradiation of the head and neck district after 15 minutes.} 
\label{tableH10head}
\begin{tabular}{c|c|c|c|c}
 Radius (cm)  &  H* 1h irr. ($\mu$Sv/h)  & Rel. err. & H* 2h irr. ($\mu$Sv/h)  & Rel. err.  \\ \hline
30-40 & 3.21$\cdot 10^1$ & 1.5$\cdot 10^{-2}$ & 5.33$\cdot 10^1$  & 1.5$\cdot 10^{-2}$ \\ \hline
40-50 & 2.35$\cdot 10^1$ & 1.5$\cdot 10^{-2}$ & 3.90$\cdot 10^1$ & 1.5$\cdot 10^{-2}$\\ \hline
50-60 & 1.83$\cdot 10^1$ & 1.5$\cdot 10^{-2}$ & 3.03$\cdot 10^1$ & 1.5$\cdot 10^{-2}$ \\ \hline
60-70 & 1.47$\cdot 10^1$ & 1.5$\cdot 10^{-2}$ & 2.45$\cdot 10^1$ & 1.5$\cdot 10^{-2}$ \\ \hline
70-80 & 1.22$\cdot 10^1$ & 1.5$\cdot 10^{-2}$ & 2.03$\cdot 10^1$ & 1.5$\cdot 10^{-2}$ \\ \hline
80-90 & 1.03$\cdot 10^1$ & 1.5$\cdot 10^{-2}$ & 1.71$\cdot 10^1$ & 1.5$\cdot 10^{-2}$ \\ \hline
90-100 & 8.84 & 1.5$\cdot 10^{-2}$ & 1.47$\cdot 10^1$ & 2.0$\cdot 10^{-2}$ \\ \hline
100-110 & 7.65 & 1.5$\cdot 10^{-2}$ & 1.27$\cdot 10^1$ & 2.0$\cdot 10^{-2}$\\ \hline
110-120 & 6.70 & 2.0$\cdot 10^{-2}$ & 1.11$\cdot 10^1$ & 2.0$\cdot 10^{-2}$\\ 
\end{tabular}
\end{table}

The values at distances between 30-40 cm are meaningful, as they represent the dose that medical staff would absorb assisting the patient.
The dose at 100-110 cm is also interesting because that represents the typical distance between individuals.

\subsubsection{Irradiation of the thoracic district}
For both the irradiation times for the thorax district, the three most activated regions are:
\begin{itemize}
    \item N° 10700, trunk muscles;
    \item N° 11700, trunk residual soft tissues;
    \item N° 9500, liver.
\end{itemize}

In this case the liver is one of the most activated organs. 
External shielding might be used to lower activation of the out-of-field organs, although activation is also due to internal neutron scattering and diffusion.

Tables  \ref{table3.2}, 
\ref{table3.4}, 
\ref{table3.6},
list, for these regions, the three most relevant isotopes, in terms of relative percentage of total activation, for 1 hour irradiation at the end of treatment (+0 s) and at 15 minutes after the shutdown (+15 m).

\begin{table}[H]
\centering
\caption{Activation of isotopes in Region 10700 (trunk muscles), for 1 h Irradiation at the beam shutdown (+0 s) and after 15 minutes (+15 m). Statistical uncertainty below  0.5\%.} 
\label{table3.2}
\begin{tabular}{c|c|c|c|c|c}
 Isotope      & Activity(Bq)  & Rel activity &  Activity(Bq) & Rel activity &  Half-life (s) \\ 
  & +0 s & +0 s & +15 m & +15 m & \\ \hline
 $^{24^m}$Na & 1.690 $\cdot 10^7$ & 82.75\%  & - & - & 2.018 $\cdot 10^{-2}$ \\ \hline
  $^{38}$Cl & 1.935 $\cdot 10^6$ & 9.47\%  & 1.464 $\cdot 10^6$ & 48.32\% & 2.234 $\cdot 10^3$ \\ \hline
  $^{24}$Na & 1.017 $\cdot 10^6$  & 4.98\%  & 1.005 $\cdot 10^6$ & 33.19\%  &   5.398 $\cdot 10^4$ \\ \hline
  $^{42}$K & - & - & 5.333 $\cdot 10^5$  & 17.60\%  &  4.450 $\cdot 10^4$ 
\end{tabular}
\end{table}


\begin{table}[H]
\centering
\caption{Activation of isotopes in Region 11700 (trunk soft tissue), for 1 h Irradiation at th beam shutdown (+0 s) and after 15 minutes (+15 m). Statistical uncertainty below  0.5\%.} 
\label{table3.4}
\begin{tabular}{c|c|c|c|c|c}
 Isotope      & Activity(Bq)  & Rel activity &  Activity(Bq) & Rel activity &  Half-life (s) \\ 
  & +0 s & +0 s & +15 m & +15 m & \\ \hline
 $^{24^m}$Na & 1.603 $\cdot 10^7$ & 85.11\%   & - & - & 2.018 $\cdot 10^{-2}$ \\ \hline
  $^{38}$Cl & 1.826 $\cdot 10^6$ & 9.70\%  & 1.381 $\cdot 10^6$ & 58.81\% & 2.234 $\cdot 10^3$ \\ \hline
  $^{24}$Na & 9.647 $\cdot 10^5$  & 5.12\%  & 9.562 $\cdot 10^5$ & 40.60\%   &  5.398 $\cdot 10^4$ \\ \hline
   $^{32}$P & - & - & 9.962 $\cdot 10^3$  & 0.42\%   &  1.232 $\cdot 10^6$
\end{tabular}
\end{table}


\begin{table}[H]
\centering
\caption{Activation of isotopes in Region 9500 (liver), for 1 h Irradiation at the beam shutdown (+0 s) and after 15 minutes (+15 m). Statistical uncertainty below  0.5\%.} 
\label{table3.6}
\begin{tabular}{c|c|c|c|c|c}
 Isotope      & Activity(Bq)  & Rel activity &  Activity(Bq) & Rel activity &  Half-life (s) \\ 
  & +0 s & +0 s & +15 m & +15 m & \\ \hline
 $^{24^m}$Na & 5.704 $\cdot 10^6$ & 84.14\% & - & -  & 2.018 $\cdot 10^{-2}$ \\ \hline
  $^{38}$Cl & 6.547 $\cdot 10^5$ & 9.66\%  & 4.952 $\cdot 10^5$ & 54.63\% &   2.234 $\cdot 10^3$ \\ \hline
  $^{24}$Na & 3.432 $\cdot 10^5$  & 5.07\%   & 3.398 $\cdot 10^5$ & 37.43\% &  5.398 $\cdot 10^4$ \\ \hline
$^{42}$K & - & - & 6.691 $\cdot 10^4$  & 7.38\%   & 4.450 $\cdot 10^4$
\end{tabular}
\end{table}


The same isotopes are activated as in the irradiation of the head and neck district, thus, the same observations hold.
In addition, as listed in Table \ref{table3.4}, $^{32}$P is present with an activity equal to 0.42\% of the total, with half-life 1.232 $\cdot 10^6$ s.
This means that, after some days, the activity of the short half-life isotopes will be negligible and the main contribution will come from $^{32}$P. However, given its low activity, its contribution would not be a strong constraint.

\begin{figure}
\centering
 [a]
{\includegraphics[width=5.9cm]{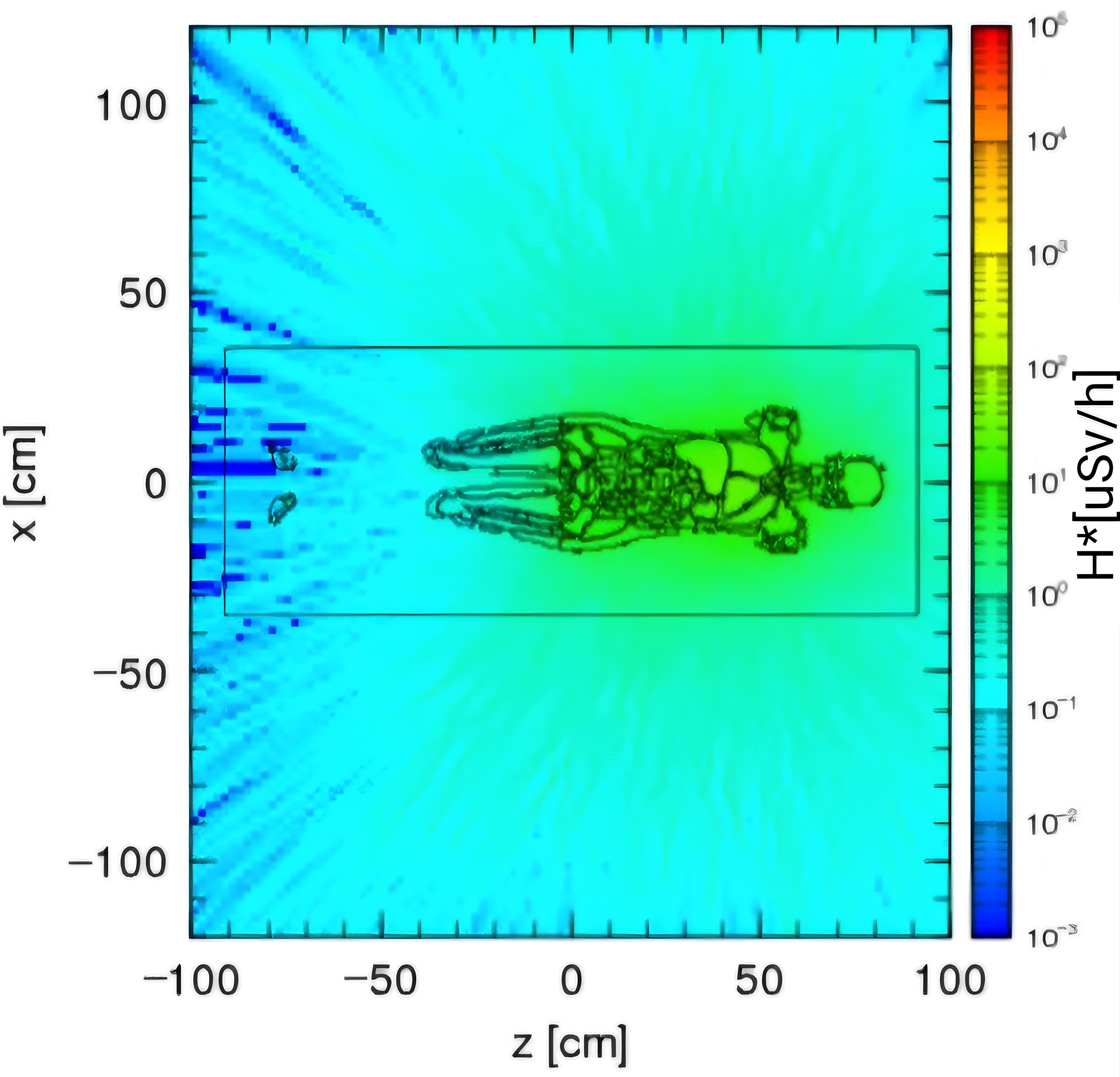}}
\hspace{5mm}
 [b]
{\includegraphics[width=5.7cm]{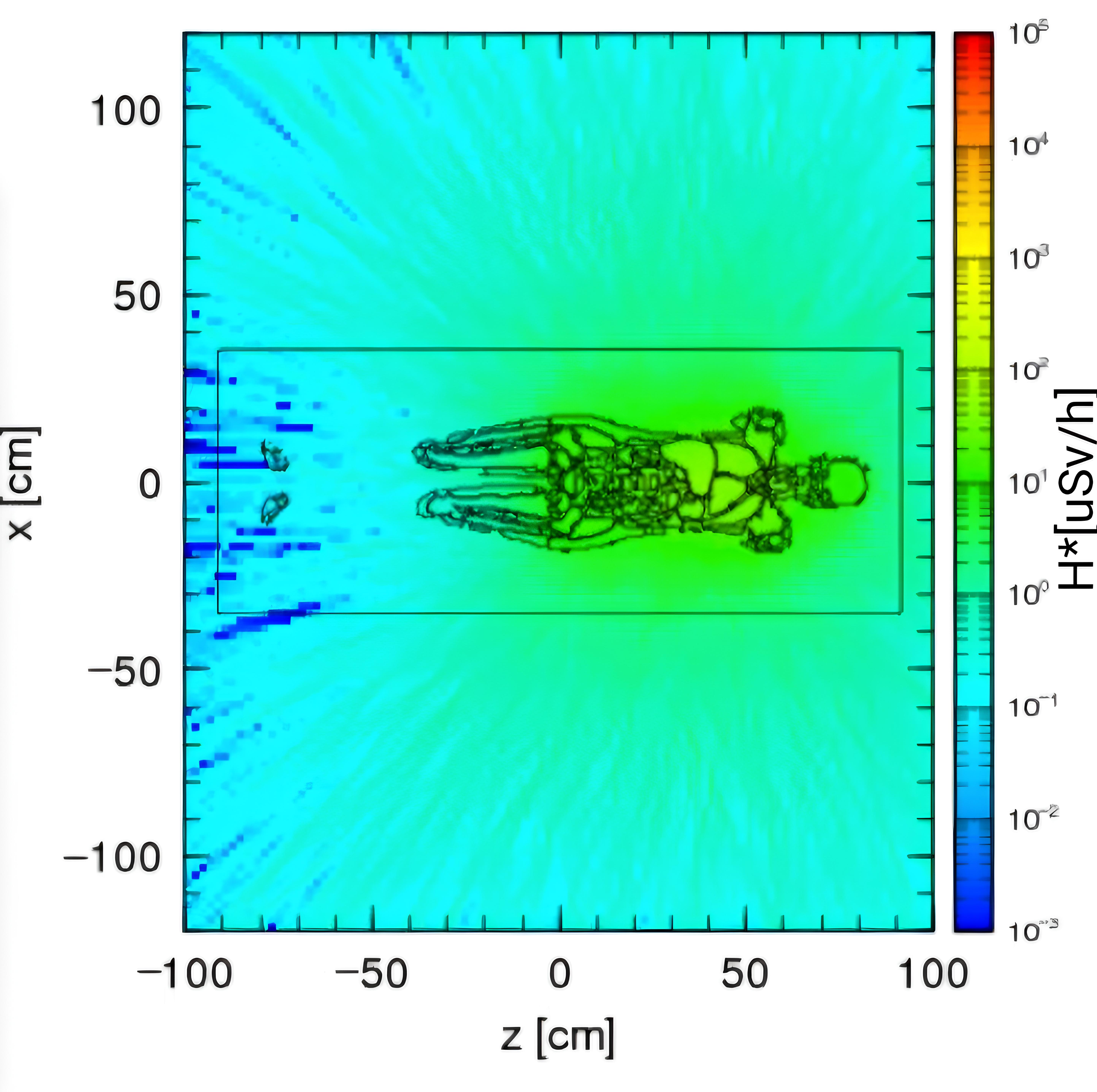}}   
\caption{H* maps at 15 minutes after 1 hour [a] and 2 hours [b] of thorax irradiation.}
\label{image11}
\end{figure}

Figure \ref{image11} shows the H* rate maps for the thorax at 15 minutes after 1 hour and 2 hours of irradiation time. As expected, the highest contribution to the dose comes from the irradiated district.

Table \ref{tableH10thorax} reports the sum of the H* rate values over the whole cylindrical mesh height for each ring of the mesh.

\begin{table}[H]
\centering
\caption{H* rate produced by the patient activation due to irradiation of the thoracic district at 15 minutes after the beam shutdown.} 
\label{tableH10thorax}
\begin{tabular}{c|c|c|c|c}
 Radius (cm)      &  H* 1h irr. ($\mu$Sv/h)  & Rel. err. & H* 2h irr. ($\mu$Sv/h)  & Rel. err.  \\ \hline
30-40 & 3.27$\cdot 10^1$ & 1.0$\cdot 10^{-2}$ & 5.65$\cdot 10^1$ & 1.0$\cdot 10^{-2}$ \\ \hline
40-50 & 2.31$\cdot 10^1$ & 1.0$\cdot 10^{-2}$ & 3.99$\cdot 10^1$ & 1.5$\cdot 10^{-2}$ \\ \hline
50-60 & 1.74$\cdot 10^1$ & 1.5$\cdot 10^{-2}$ & 3.00$\cdot 10^1$ & 1.5$\cdot 10^{-2}$ \\ \hline
60-70 & 1.36$\cdot 10^1$ & 1.5$\cdot 10^{-2}$ & 2.35$\cdot 10^1$ & 1.5$\cdot 10^{-2}$ \\ \hline
70-80 & 1.10$\cdot 10^1$ & 1.5$\cdot 10^{-2}$ & 1.89$\cdot 10^1$ & 1.5$\cdot 10^{-2}$ \\ \hline
80-90 & 9.04 & 1.5$\cdot 10^{-2}$ & 1.56$\cdot 10^1$ & 1.5$\cdot 10^{-2}$ \\ \hline
90-100 & 7.58 & 1.5$\cdot 10^{-2}$ & 1.31$\cdot 10^1$ & 1.5$\cdot 10^{-2}$ \\ \hline
100-110 & 6.44 & 1.5$\cdot 10^{-2}$ & 1.11$\cdot 10^1$ & 1.5$\cdot 10^{-2}$ \\ \hline
110-120 & 5.54 & 1.5$\cdot 10^{-2}$ & 9.56 & 2.0$\cdot 10^{-2}$ \\ 

\end{tabular}
\end{table}

\subsubsection{Irradiation of the lower limbs district}
When irradiating the lower limbs, the three most activated regions are:
\begin{itemize}
    \item N° 10900, legs muscles;
    \item N° 11900, legs residual soft tissues;
    \item N° 3400, tibia, fibulae and patellae cortical bone.
\end{itemize}

Tables \ref{table4.2}, 
\ref{table4.4}, 
\ref{table4.6},
list the three most relevant isotopes, in terms of relative percentage of total activation, for 1 hour irradiation at the beam shutdown (+0 s) and after 15 minutes (+15 m).

\begin{table}[H]
\centering
\caption{Activation of isotopes in Region 10900 (leg muscles), for 1 h Irradiation at the beam shutdown (+0 s), and after 15 minutes (+15 m). Statistical uncertainty below  0.5\%.} 
\label{table4.2}
\begin{tabular}{c|c|c|c|c|c}
 Isotope      & Activity(Bq)  & Rel activity &  Activity(Bq) & Rel activity &  Half-life (s) \\ 
  & +0 s & +0 s & +15 m & +15 m & \\ \hline
 $^{24^m}$Na & 8.924 $\cdot 10^6$ & 82.75\% & - & -    & 2.018 $\cdot 10^{-2}$ \\ \hline
  $^{38}$Cl & 1.022 $\cdot 10^6$ & 9.48\% & 7.731 $\cdot 10^5$ & 48.34\%  & 2.234 $\cdot 10^3$ \\ \hline
  $^{24}$Na & 5.369 $\cdot 10^5$  & 4.98\% & 5.308 $\cdot 10^5$ & 33.19\%  &   5.398 $\cdot 10^4$ \\ \hline
    $^{42}$K & - & - & 2.807 $\cdot 10^5$  & 17.55\%  &  4.450 $\cdot 10^4$
\end{tabular}
\end{table}


\begin{table}[H]
\centering
\caption{Activation of isotopes in Region 11900 (legs soft tissue), for 1 h Irradiation at the beam shutdown (+0 s) and after 15 minutes (+15 m). Statistical uncertainty below  0.5\%.} 
\label{table4.4}
\begin{tabular}{c|c|c|c|c|c}
 Isotope      & Activity(Bq)  & Rel activity &  Activity(Bq) & Rel activity &  Half-life (s) \\ 
  & +0 s & +0 s & +15 m & +15 m & \\ \hline
 $^{24^m}$Na & 6.332 $\cdot 10^6$ & 85.13\% & - & -  & 2.018 $\cdot 10^{-2}$ \\ \hline
  $^{38}$Cl & 7.193 $\cdot 10^5$ & 9.67\% & 5.440 $\cdot 10^5$ & 58.74\%  & 2.234 $\cdot 10^3$ \\ \hline
  $^{24}$Na & 3.810 $\cdot 10^5$  & 5.12\%  & 3.766 $\cdot 10^5$ & 40.66\%  &  5.398 $\cdot 10^4$ \\ \hline
 $^{32}$P & - & - &  3.925 $\cdot 10^3$  & 0.42\%  &  1.232 $\cdot 10^6$  
\end{tabular}
\end{table}


\begin{table}[H]
\centering
\caption{Activation of isotopes in Region 3400 (cortical bone), after 1 h irradiation at the beam shutdown (+0 s) and after 15 minutes (+15 m). Statistical uncertainty below  0.5\%.} 
\label{table4.6}
\begin{tabular}{c|c|c|c|c|c}
 Isotope      & Activity(Bq)  & Rel activity &  Activity(Bq) & Rel activity &  Half-life (s) \\ 
  & +0 s & +0 s & +15 m & +15 m & \\ \hline
 $^{24^m}$Na & 3.287 $\cdot 10^6$ & 72.24\% & - & -  & 2.018 $\cdot 10^{-2}$ \\ \hline
  $^{49}$Ca  & 6.795 $\cdot 10^5$ & 14.94\% & 2.062 $\cdot 10^5$ & 26.17\% &  5.231 $\cdot 10^2$ \\ \hline
  $^{49}$Sc & 2.956 $\cdot 10^5$  & 6.50\% & 3.113 $\cdot 10^5$ & 39.51\%   &  3.431 $\cdot 10^3$\\ \hline
  $^{24}$Na & - & - & 1.955 $\cdot 10^5$  & 24.81\%   & 5.398 $\cdot 10^4$  
\end{tabular}
\end{table}


The relative activity for the regions 10900 and 11900 is almost equal, respectively, to the one for the cells 10700 and 11700 in the case of thorax irradiation due to the fact that these cells are composed of the same material. The absolute activity instead is different due to the different shape and quantity of irradiated material.

\begin{figure}[H]
\centering
 [a]
{\includegraphics[width=5.8cm]{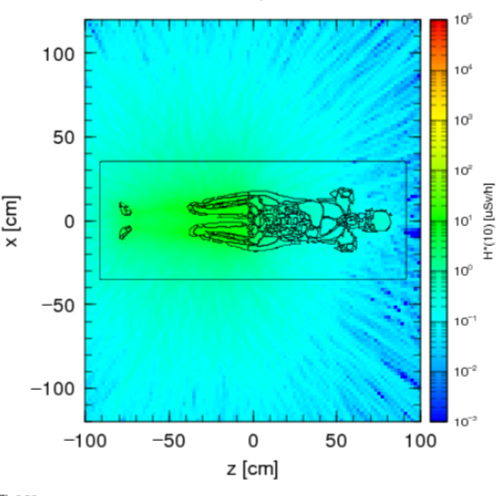}}
\hspace{5mm}
 [b]
{\includegraphics[width=5.9cm]{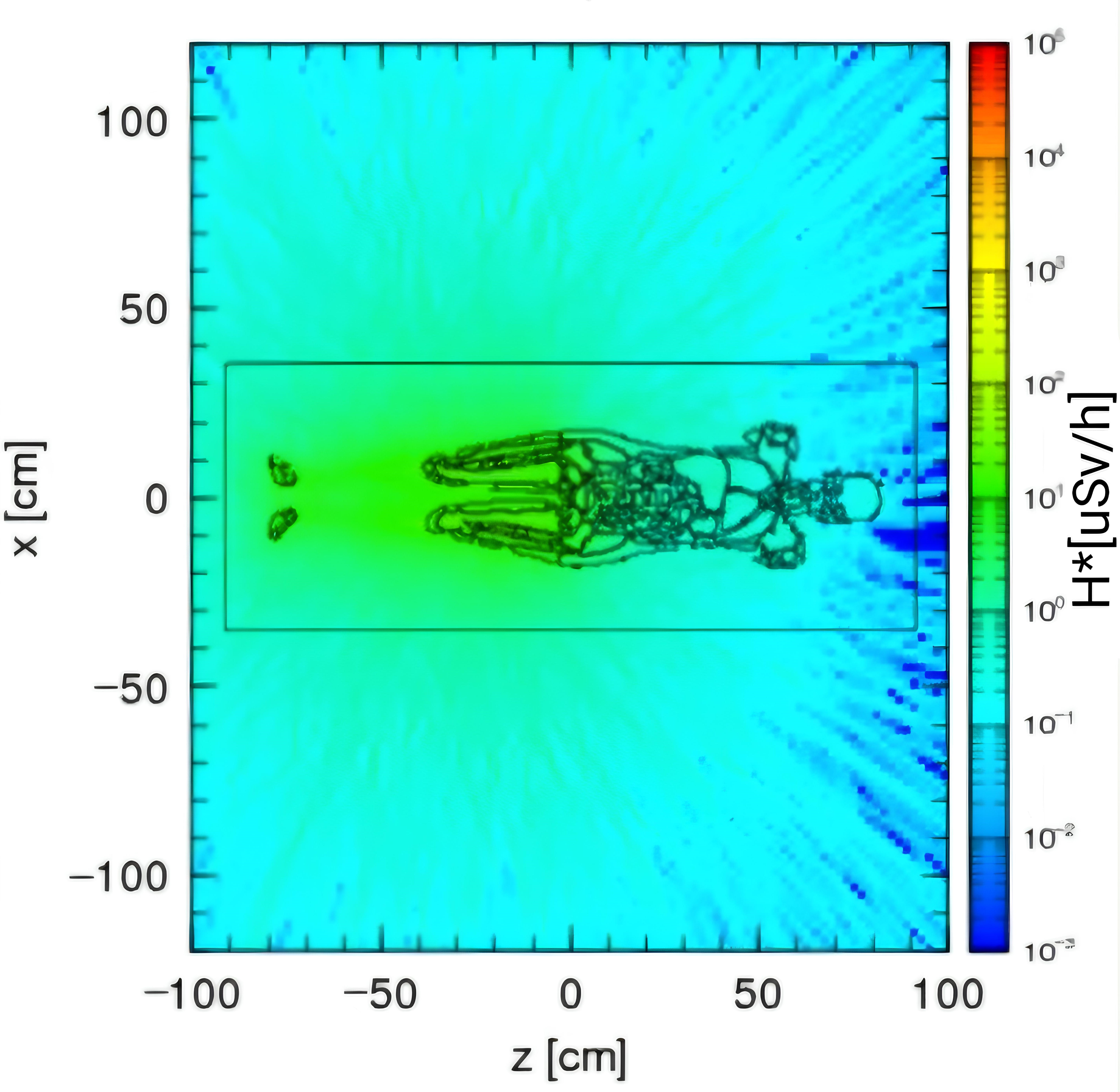}}   
\caption{H* maps at 15 minutes after 1 hour [a] and 2 hours [b] of lower limbs irradiation.}
\label{image13}
\end{figure}

Figure \ref{image13} represents the H* map for the lower limbs district irradiation at 15 minutes after 1 hour and 2 hours irradiation.

Table \ref{tableH10legs} reports the values of the equivalent ambient dose rate produced by the patient activation due to the lower limbs district irradiation in a radius of 120 cm.

\begin{table}[H]
\caption{H* rate produced by the patient activation due to legs irradiation after 15 minutes} 
\label{tableH10legs}
\begin{tabular}{c|c|c|c|c}
 Radius (cm)      &  H* 1h irr. ($\mu$Sv/h)  & Rel. err. & H* 2h irr. ($\mu$Sv/h)  & Rel. err.  \\ \hline
30-40 & 1.84$\cdot 10^1$ & 1.0$\cdot 10^{-2}$ & 3.17$\cdot 10^1$ & 1.0$\cdot 10^{-2}$ \\ \hline
40-50 & 1.31$\cdot 10^1$ & 1.0$\cdot 10^{-2}$ & 2.26$\cdot 10^1$ & 1.5$\cdot 10^{-2}$ \\ \hline
50-60 & 9.93 & 1.5$\cdot 10^{-2}$ & 1.71$\cdot 10^1$ & 1.5$\cdot 10^{-2}$ \\ \hline
60-70 & 7.80 & 1.5$\cdot 10^{-2}$ & 1.34$\cdot 10^1$ & 1.5$\cdot 10^{-2}$ \\ \hline
70-80 & 6.30 & 1.5$\cdot 10^{-2}$ & 1.08$\cdot 10^1$ & 1.5$\cdot 10^{-2}$ \\ \hline
80-90 & 5.19 & 1.5$\cdot 10^{-2}$ & 8.93 & 1.5$\cdot 10^{-2}$ \\ \hline
90-100 & 4.33 & 1.5$\cdot 10^{-2}$ & 7.46 & 1.5$\cdot 10^{-2}$ \\ \hline
100-110 & 3.68 & 1.5$\cdot 10^{-2}$ & 6.33 & 1.5$\cdot 10^{-2}$ \\ \hline
110-120 & 3.16 & 1.5$\cdot 10^{-2}$ & 5.45 & 1.5$\cdot 10^{-2}$ \\ 

\end{tabular}
\end{table}

\subsection{Comparison with $^{131}$I}
Knowing the distribution of H* rate due to the patient activation at each different position, it is now possible to evaluate the impact of these results by comparison to the values obtained for the 600 MBq $^{131}$I administration.

Table \ref{tablecomp} reports the values of the simulations at the two meaningful distances mentioned above: 30-40 cm 
and 100-110 cm.
To maintain these considerations as conservative as possible, the comparison was carried out with the 2 hours irradiation case, despite the irradiation of patients is shorter than 1 hour in current clinical applications with accelerators \cite{suzuki2023initial}, \cite{hirose2021boron}, \cite{igaki2022scalp}.

\begin{table}[H]
\centering
\caption{H* rate comparison between the ambient dose due to iodine administration and 2-hour BNCT irradiation after 15 minutes. Statistical uncertainty lower than 2\%.} 
\begin{tabular}{c|c|c|c|c}
 Radius (cm)  & $^{131}$I ($\mu$Sv/h) &  Head ($\mu$Sv/h)  &  Thorax ($\mu$Sv/h) &   Legs ($\mu$Sv/h) \\ \hline
30-40 & 1.15$\cdot 10^3$ & 5.33$\cdot 10^{1}$ & 5.65$\cdot 10^1$ & 3.17$\cdot 10^{1}$ \\ \hline
100-110 & 2.49$ \cdot 10^2$ & 1.27$\cdot 10^{1}$ & 1.11 $\cdot 10^1$ & 6.33$ \cdot 10^0$\\ 
\end{tabular}
\label{tablecomp}
\end{table}

\begin{figure}
\centering
 [a]
{\includegraphics[width=5.54cm]{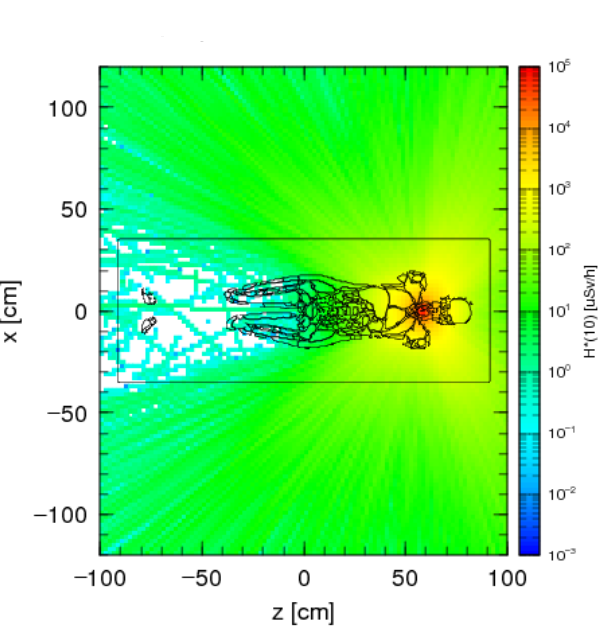}}
\hspace{5mm}
 [b]
{\includegraphics[width=5.5cm]{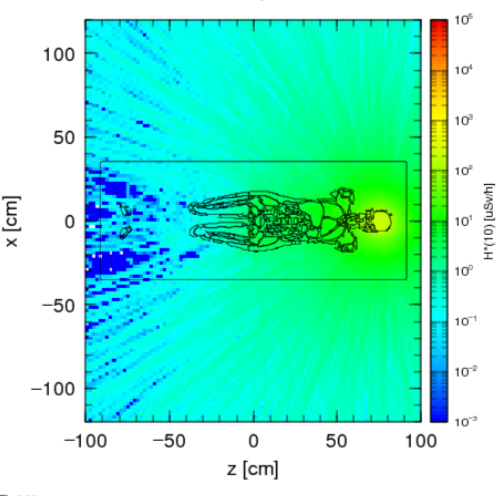}}   

\centering
 [c]
{\includegraphics[width=5.3cm]{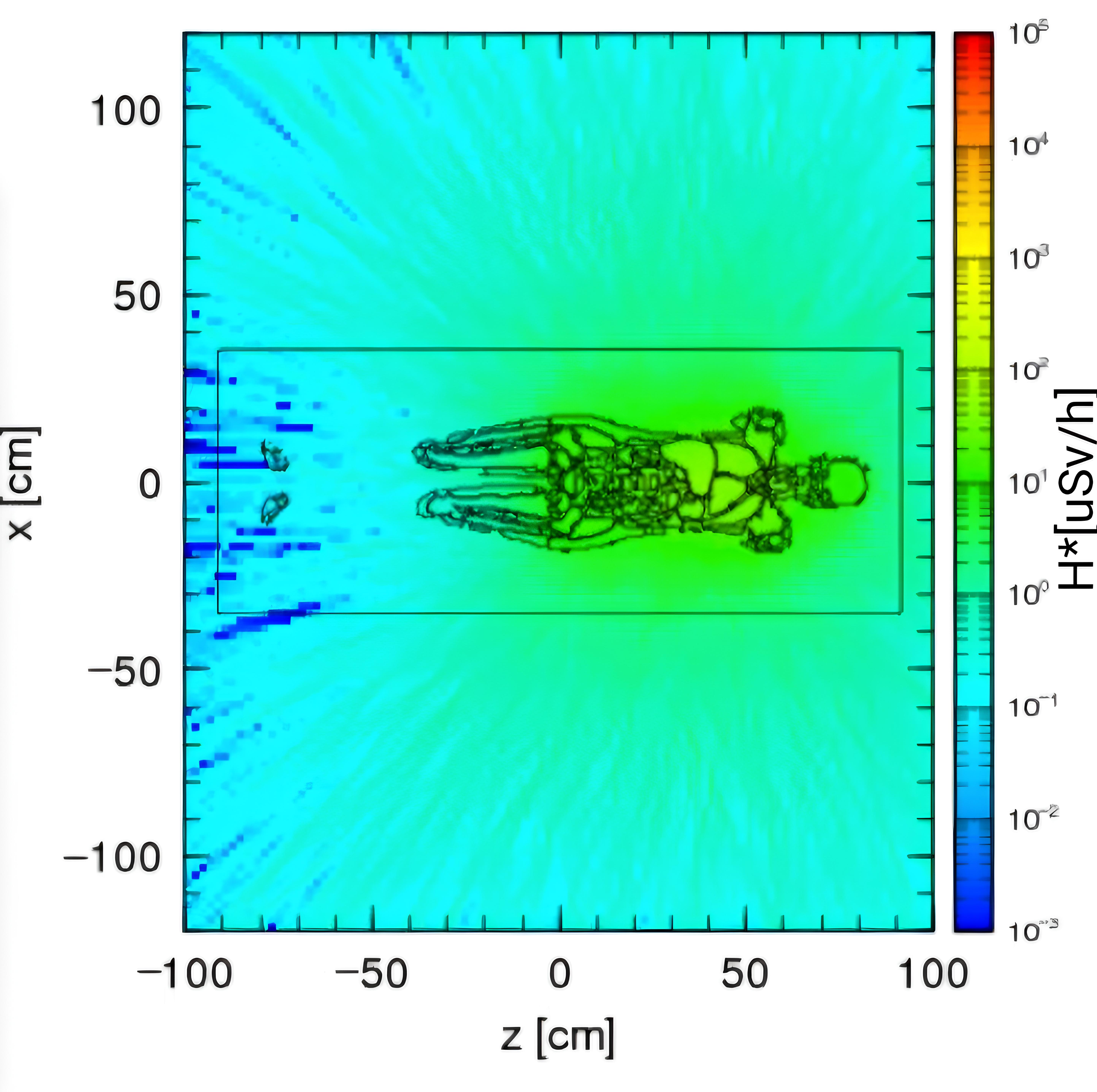}}
\hspace{5mm}
 [d]
{\includegraphics[width=5.5cm]{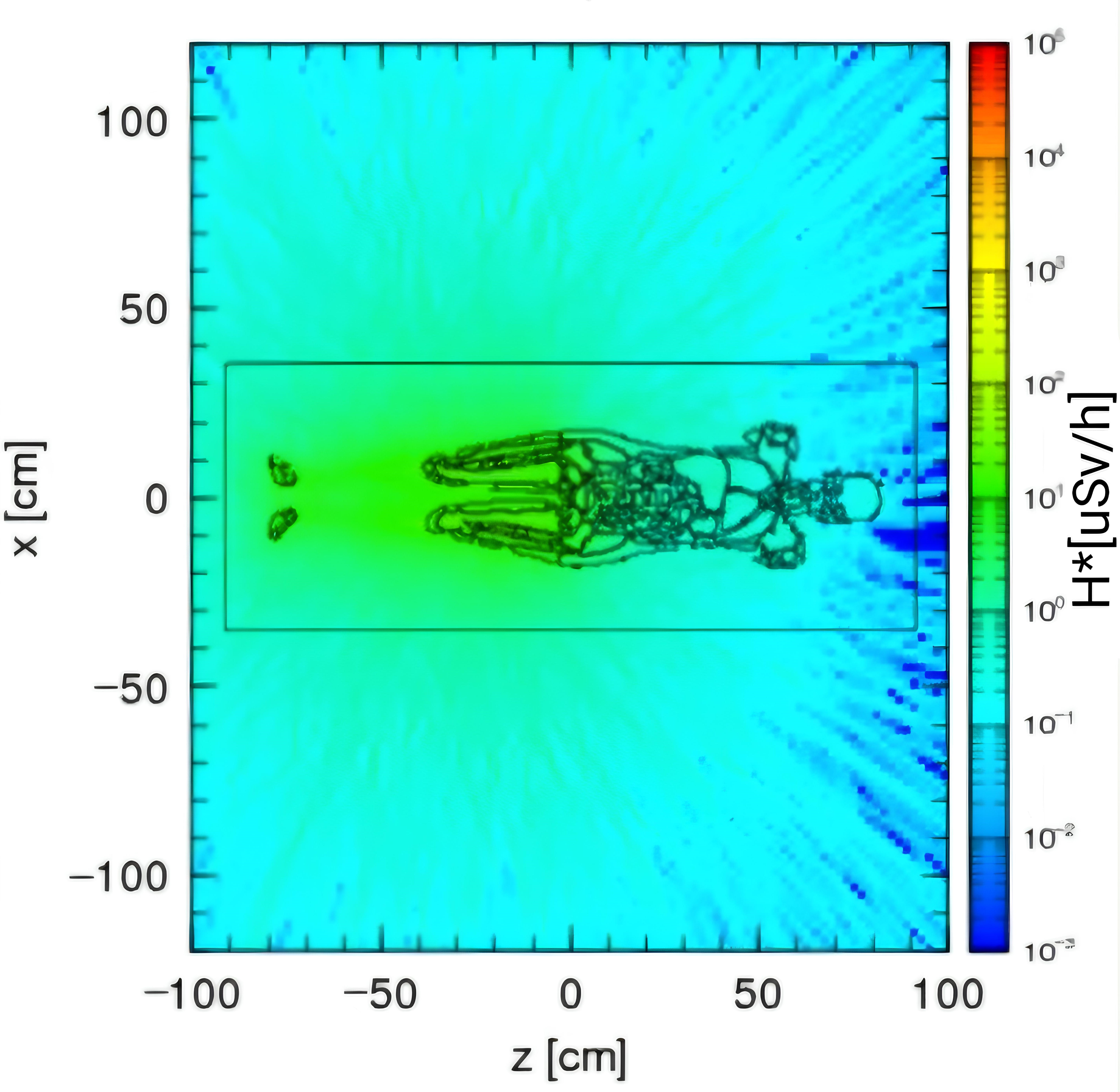}}   
\caption{H* maps due to $^{131}$I administration [a] and due to 2 hours irradiation activation (after 15 minutes) for head [b], thorax [c], legs [d].}
\label{imagecomp2}
\end{figure}

Figure \ref{imagecomp2} shows the H* rate maps due to 600 MBq of iodine-131 in the thyroid, and the H* rate at 15 minutes after the end of irradiation for the treatment of the three districts described above.

From Table \ref{tablecomp} and Figure \ref{imagecomp2} it is evident that the $^{131}$I (concentrated in the thyroid) produces a higher ambient dose with respect to the neutron irradiation even in the most conservative assumptions. The difference is almost of two order of magnitude at 15 minutes after 2 hours irradiation. This means that, from the radiation protection point of view, and adopting this criterion adapted from nuclear medicine in current Italian regulation, the patient could be discharged already after 15 minutes post irradiation. 
In BNCT centers treating patients, after the irradiation patients are kept in a dedicated room for about one hour to observe the general conditions and to rest. 
Thus, at the moment of patients discharge, the ambient dose will be even lower and it will not constitute a radiation protection issue. This prediction is well confirmed by current clinical experience in accelerator-based BNCT facility, for example in Xiamen, where patients are discharged after activation monitoring (personal communication).

\subsection{Activation of Urine}
At the end of the irradiation, for both 1 hour and 2 hours irradiation time, the highest contribution to the activity in urine ( $\sim$ 73\% at 1 hour and $\sim$ 80\% at 2 hours for all the irradiation positions) is given by the $^{24^m}$Na.
The half-life of this isotope is  2.018 $\cdot 10^{-2}$ s, hence its activity is  significantly reduced already one second after the beam shutdown.

Table \ref{tableH_urine}
reports the most activated isotopes in urine for each position at 15 minutes after 1 hour and 2 hours of irradiation. This time interval was selected as the reference for this evaluation, as it corresponds to the minimum patient's stay in the treatment room after irradiation. This interval also marks the earliest point at which urine can be collected in the containers.
In addition to the results obtained, it is noteworthy to highlight the activity of $^{35}$S, which is relevant due to its long half-life.

\begin{table}[H]
\centering
\caption{Most relevant isotopes activated in the irradiation of the three districts. Statistical uncertainty is below 4\%.}
\label{tableH_urine}
\begin{tabular}{c|c|c|c}
\multicolumn{4}{c}{Head and Neck}\\ 
\hline
& 1h activity (Bq)  & 2h activity (Bq) &  Half-life (s) \\ \hline
$^{38}$Cl & 6.664 $\cdot 10^2 $ & 8.845 $\cdot 10^2 $ &  2.234 $\cdot 10^3$ \\ \hline
$^{24}$Na & 3.049 $\cdot 10^2 $ & 5.460 $\cdot 10^2 $ & 5.398 $\cdot 10^4$  \\ \hline
$^{42}$K  & 2.006 $\cdot 10^1$ & 3.902 $\cdot 10^1$ & 4.450 $\cdot 10^4$ \\ \hline
$^{35}$S  & 1.545 & 3.094 & 7.561 $\cdot 10^6$\\ \hline
\multicolumn{4}{c}{Thorax}\\
\hline
& 1h activity (Bq)  & 2h activity (Bq) &  Half-life (s) \\ \hline
$^{38}$Cl & 2.602 $\cdot 10^3 $ & 3.454 $\cdot 10^3 $ &  2.234 $\cdot 10^3$ \\ \hline
$^{24}$Na & 1.190 $\cdot 10^3 $ & 2.326 $\cdot 10^3 $ & 5.398 $\cdot 10^4$  \\ \hline
$^{42}$K & 7.800 $\cdot 10^1$ & 1.517 $\cdot 10^2$ & 4.450 $\cdot 10^4$ \\ \hline
$^{35}$S & 6.045 & 1.209 $\cdot 10^1$ & 7.561 $\cdot 10^6$\\ \hline
\multicolumn{4}{c}{Lower limbs}\\ \hline
& 1h activity (Bq)  & 2h activity (Bq) &  Half-life (s) \\ \hline
$^{38}$Cl & 6.050 $\cdot 10^3 $ & 8.030 $\cdot 10^3 $ &  2.234 $\cdot 10^3$ \\ \hline
$^{24}$Na & 2.764 $\cdot 10^3 $ & 5.403$\cdot 10^3 $ & 5.398 $\cdot 10^4$  \\ \hline
$^{42}$K & 1.816 $\cdot 10^2$ & 3.533 $\cdot 10^2$ & 4.450 $\cdot 10^4$ \\ \hline
$^{35}$S & 1.401 $\cdot 10^1$ & 2.802 $\cdot 10^1$ & 7.561 $\cdot 10^6$
\end{tabular}
\end{table}



Among the listed isotopes also $^{40}$K is relevant, with an activity ranging from some Bq (in the irradiation of the head and neck region) to few hundreds of Bq (in the irradiation of the lower limbs). This isotope is interesting because of its long half-life ($10^9$ years). 
The conclusion of this analysis is that, after about ten days, the main contributions will come from the $^{35}$S and $^{40}$K.


\begin{table}
\centering
\begin{tabular}{ccc}
\multicolumn{1}{c|}{} & \multicolumn{1}{c|}{Head 1h} & Head 2h \\ \hline
\multicolumn{1}{c|}{Activity (Bq)} & \multicolumn{1}{c|}{1.01 $\cdot 10^3$ (
3\%)} & 1.54$\cdot 10^3$ (3\%) \\ \hline
\multicolumn{1}{c|}{Specific activity (Bq/g)} & \multicolumn{1}{c|}{5.23} & 7.99\\
 &  &  \\
\multicolumn{1}{c|}{} & \multicolumn{1}{c|}{Thorax 1h} & Thorax 2h \\ \hline
\multicolumn{1}{c|}{Activity (Bq)} & \multicolumn{1}{c|}{3.89$\cdot 10^3$ (3\%)} & 5.96$\cdot 10^3$(3\%) \\ \hline
\multicolumn{1}{c|}{Specific activity (Bq/g)} & \multicolumn{1}{c|}{20.24} & 31.01 \\
\multicolumn{1}{l}{} &  &  \\
\multicolumn{1}{l|}{} & \multicolumn{1}{c|}{Legs 1h} & Legs 2h \\ \hline
\multicolumn{1}{c|}{Activity (Bq)} & \multicolumn{1}{c|}{ 9.03$\cdot 10^3$ (2\%)} &  1.38$\cdot 10^4$(2\%) \\ \hline
\multicolumn{1}{c|}{Specific activity (Bq/g)} & \multicolumn{1}{c|}{46.95} & 71.98
\end{tabular}
\caption{Urine activity at 15 minutes after the irradiation for each position. In brackets is reported the statistical uncertainty in percentage.}
\label{urineact}
\end{table}

Table \ref{urineact} shows the total and the specific (per unit gram) urine activity for the two irradiation times and for each treatment position at 15 minutes after the irradiation.
It is clear that the irradiation of the lower limbs district causes a higher activation of the urine because of the higher neutron flux in the bladder.

\section{Conclusions}\label{concl}
This work has explored the relevance of the patient activation for radiation protection purposes in BNCT, assuming conservative irradiation times. 

This study used the computational adult-patient model from ICRP-145 to benchmark two Monte Carlo transport codes commonly used in medical physics. PHITS was identified as the most suitable choice due to its faster calculation times and its compatibility with the DCHAIN-SP software, enabling the determination of radiation sources from isotope activation.

Being BNCT a new type of radiation treatment that will be available in Italy only in the next few years, at the moment there is no specific regulation about patient discharge after the treatment. This study introduces a possible criterion for assessing the relevance of patient activation following BNCT. This criterion compares the ambient dose generated by the irradiated patient to the one of a patient administered with 600 MBq of iodine-131 for thyroid treatment. This value was chosen since, according to the actual Italian legislation on radiation protection, this is the threshold under which the hospitalization of treated patients is not mandatory. Our findings demonstrate that, even under an extremely conservative irradiation time of 2 hours, patient activation does not pose a significant radiological concern based on the proposed figure of merit. These results suggest the potential for early patient discharge, possibly as soon as 15 minutes after BNCT treatment.

A more comprehensive analysis will be necessary in the future to determine the applicability of this indicator across various irradiation scenarios and to establish its validity as a benchmark for operational decision-making. Anyway, this method represents a starting point for understanding the magnitude of the effect, using a reasonable comparison approach.

The simulations also suggest that the activation of certain tissues/organs, such as the liver and bladder, could be significantly reduced through the implementation of shielding to protect the patient's body during irradiation. This aspect should be included in the treatment planning phase of each patient, considering the patient's specific positioning relative to the beam port.
Moreover, it will be interesting to consider some variations in the phantom composition, for example evaluating the impact of prosthesis, such as titanium hips, in the induced activity and in the H*.

The results obtained for the urine activation indicate the need of a dedicated hot restroom. This finding shows the critical importance of comprehensive radiation protection studies during the design phase of any BNCT facility, as these studies directly influence the overall design and operational management of the center.  

\section*{Acknowledgements}
This work was funded by the National Plan for NRRP Complementary Investments (PNC, established with the decree-law 6 May 2021, n. 59, converted by law n. 101 of 2021) in the call for the funding of research initiatives for technologies and innovative trajectories in the health and care sectors (Directorial Decree n. 931 of 06-06-2022) - \textbf{project n. PNC0000003 - AdvaNced Technologies for Human-centrEd Medicine (project acronym: ANTHEM)}. This work reflects only the authors’ views and opinions, neither the Ministry for University and Research nor the European Commission can be considered responsible for them. 

 \bibliographystyle{elsarticle-num} 
 \bibliography{biblio-new}

\begin{thebibliography}{10}
\expandafter\ifx\csname url\endcsname\relax
  \def\url#1{\texttt{#1}}\fi
\expandafter\ifx\csname urlprefix\endcsname\relax\def\urlprefix{URL }\fi
\expandafter\ifx\csname href\endcsname\relax
  \def\href#1#2{#2} \def\path#1{#1}\fi

\bibitem{IAEA-TecDoc2023}
AA.VV., Advances in Boron Neutron Capture Therapy, TECDOC Series, INTERNATIONAL ATOMIC ENERGY AGENCY, Vienna, 2023.

\bibitem{pisent2014munes}
A.~Pisent, E.~Fagotti, P.~Colautti, Munes a compact neutron source for bnct and radioactive wastes characterization, in: Proc. Linear Accelerator Conf. LINAC, 2014, pp. 261--263.

\bibitem{postuma2021novel}
I.~Postuma, S.~Gonz{\'a}lez, M.~S. Herrera, L.~Provenzano, M.~Ferrarini, C.~Magni, N.~Protti, S.~Fatemi, V.~Vercesi, G.~Battistoni, et~al., A novel approach to design and evaluate bnct neutron beams combining physical, radiobiological, and dosimetric figures of merit, Biology 10~(3) (2021) 174.

\bibitem{ESPOSITO}
J.~Esposito, P.~Colautti, S.~Fabritsiev, A.~Gervash, R.~Giniyatulin, V.~Lomasov, A.~Makhankov, I.~Mazul, A.~Pisent, A.~Pokrovsky, M.~Rumyantsev, V.~Tanchuk, L.~Tecchio, Be target development for the accelerator-based spes-bnct facility at infn legnaro, Applied Radiation and Isotopes 67~(7, Supplement) (2009) S270--S273.

\bibitem{icrp145}
C.~H. Kim, Y.~Yeom, N.~Petoussi-Hen{\ss}, M.~Zankl, W.~E. Bolch, C.~Lee, C.~Choi, T.~T. Nguyen, K.~Eckerman, H.~Kim, et~al., Icrp publication 145: adult mesh-type reference computational phantoms, Annals of the ICRP 49~(3) (2020) 13--201.

\bibitem{MCNP6.3}
D.~B. Pelowitz, et~al., {MCNP} -- A General Monte Carlo N-Particle Transport Code, Version 6.3, Los Alamos National Laboratory, Los Alamos, NM, lA-UR-23-10009 (2023).

\bibitem{PHITS}
T.~Sato, Y.~Iwamoto, S.~Hashimoto, T.~Ogawa, T.~Furuta, S.~ichiro Abe, T.~Kai, P.-E. Tsai, N.~Matsuda, H.~Iwase, N.~Shigyo, L.~Sihver, K.~Niita, Features of particle and heavy ion transport code system (phits) version 3.02, Journal of Nuclear Science and Technology 55~(6) (2018) 684--690.

\bibitem{RATLIFF}
H.~N. Ratliff, N.~Matsuda, S.~ichiro Abe, T.~Miura, T.~Furuta, Y.~Iwamoto, T.~Sato, Modernization of the dchain-phits activation code with new features and updated data libraries, Nuclear Instruments and Methods in Physics Research Section B: Beam Interactions with Materials and Atoms 484 (2020) 29--41.

\bibitem{ICRU95}
AA.VV., Icru report 95: Operational quantities for external radiation exposure, Journal of the ICRU 20~(7) (2020).

\bibitem{magni2022experimental}
C.~Magni, Experimental and computational studies for an accelerator-based boron neutron capture therapy clinical facility: a multidisciplinary approach - phd thesis - university of pavia (2022).

\bibitem{woodard1986composition}
H.~Woodard, D.~White, The composition of body tissues, The British journal of radiology 59~(708) (1986) 1209--1218.

\bibitem{suzuki2023initial}
S.~Suzuki, K.~Nitta, T.~Yagihashi, P.~Eide, H.~Koivunoro, N.~Sato, S.~Gotoh, S.~Shiba, M.~Omura, H.~Nagata, et~al., Initial evaluation of accelerator-based neutron source system at the shonan kamakura general hospital, Applied Radiation and Isotopes 199 (2023) 110898.

\bibitem{hirose2021boron}
K.~Hirose, A.~Konno, J.~Hiratsuka, S.~Yoshimoto, T.~Kato, K.~Ono, N.~Otsuki, J.~Hatazawa, H.~Tanaka, K.~Takayama, et~al., Boron neutron capture therapy using cyclotron-based epithermal neutron source and borofalan (10b) for recurrent or locally advanced head and neck cancer (jhn002): An open-label phase ii trial, Radiotherapy and Oncology 155 (2021) 182--187.

\bibitem{igaki2022scalp}
H.~Igaki, N.~Murakami, S.~Nakamura, N.~Yamazaki, T.~Kashihara, A.~Takahashi, K.~Namikawa, M.~Takemori, H.~Okamoto, K.~Iijima, et~al., Scalp angiosarcoma treated with linear accelerator-based boron neutron capture therapy: a report of two patients, Clinical and Translational Radiation Oncology 33 (2022) 128--133.

\end{thebibliography}





\end{document}